\documentstyle[aps,epsf,twocolumn]{revtex}

\def\be{\begin{eqnarray}} 
\def\ee{\end{eqnarray}}

\newcommand{\eq}{\begin{equation}} \newcommand{\eqx}{\end{equation}}

\newcommand{\eqn}{\begin{eqnarray}} \newcommand{\eqnx}{\end{eqnarray}}
\newcommand{\f}[2]{\frac{#1}{#2}}

\newcommand{\tr}{\mbox{\rm tr}}
\newcommand{\VV}{{\cal V}}
\newcommand{\MM}{{\cal M}}
\newcommand{\GG}{{\cal G}}
\newcommand{\DD}{{H_0}}
\newcommand{\Sg}{\Sigma}
\newcommand{\cor}[1]{\left\langle{#1}\right\rangle}
\newcommand{\One}{\mbox{\bf 1}}

\begin{document}

\draft

\title{\bf Localization Transitions from Free Random Variables}

\author{ {\bf Romuald A. Janik}$^1$, {\bf Maciej A.  Nowak}$^{1,2}$ ,
{\bf G\'{a}bor Papp}$^{3}$ and {\bf Ismail Zahed}$^4$}

\address{$^1$ Department of Physics, Jagellonian University, 30-059
Krakow, Poland.\\ $^2$ GSI, Plankstr. 1, D-64291 Darmstadt, Germany \&
\\ Institut f\"{u}r Kernphysik, TH Darmstadt, D-64289 Darmstadt, Germany
\\ $^3$GSI, Plankstr. 1, D-64291 Darmstadt, Germany \& \\ Institute for
Theoretical Physics, E\"{o}tv\"{o}s University, Budapest, Hungary\\
$^4$Department of Physics, SUNY, Stony Brook, New York 11794, USA.}
\date{\today} \maketitle

\begin{abstract}
We motivate and use the concept of free random variables for the studies of the
de-pinning transition of flux lines in superconductors as recently discussed by
Hatano and Nelson. We derive analytical conditions for the critical points 
of the eigenvalue distribution in the complex plane for all values of the
transverse magnetic field in one-dimension, in agreement with 
numerical calculations. We suggest a relation between reduced nonhermitean 
quantum mechanics in D-dimensions and nonhermitean random matrix models 
with weak nonhermiticity.

\end{abstract}

{\bf 1.\,\,\,}
Recently Hatano and Nelson~\cite{NELSON}
have shown that the de-pinning of flux lines from 
columnar defects in superconductors in D+1-dimensions, may be mapped onto the
world lines of bosons in D-dimensions. The pinning and hence localization by
the columnar defects is mapped onto an on site real randomness, and the 
de-pinning by the transverse magnetic field is mapped onto a directed
hoping, resulting into  non-hermitean quantum mechanics. While it is generally
accepted that all eigenstates are localized in one- and two-dimensions in the
presence of randomness, it is clear that the flux lines are 
de-pinned by a strong transverse magnetic field. The de-pinning
in one- and two-dimensions was studied numerically in~\cite{NELSON}.

In this letter, we would like to show that the non-hermitean tight-binding 
model discussed by Hatano and Nelson  in one-dimension can be analyzed in a
straightforward way using the concepts of free random variables. In section 2
and 3, we introduce the model and motivate the use of the addition law for 
free random variables. In section 4, we derive an explicit condition for the
end-points of the distribution of (localized) eigenvalues on the real axis for
arbitrary transverse magnetic field. In section 5, we extend our analysis 
to the complex eigenvalue plane. In section 6,
we use the uncertainty principle to show that the constant mode sector of 
this and related models may be amenable  to non-hermitean random matrix models 
with weak nonhermiticity~\cite{FYODOROV,EFETOV}. 

\vskip .2cm
{\bf 2.\,\,\,}
Following Hatano and Nelson, we consider the non-hermitean tight-binding 
Hamiltonian in second quantized form for $D=1$ 
\eqn
H=&& H_0 + \VV = 
\label{master} \\
 =&&\!\! \frac t2 \!\sum_{A=1}^N \!\!\left( 
	(e^{+h} \,c^{\dagger}_{A+1} c_{A}\!+\!e^{-h} \, 
	c^{\dagger}_{A-1} c_{A} ) + V_A \,c^{\dagger}_A c_A 
	\right) \nonumber
\eqnx
where $c_A^{\dagger}$ is a boson creation operator at site $A$. 
Throughout, the lattice spacing $a=1$. The diagonal
entries are random with elements distributed uniformly 
between $(-\Delta, \Delta)$, and
the deterministic part $H_0$ is off-diagonal with hopping strengths
$te^{\pm h}/2$ ($t<0$). Here $h$ is the typical strength of the 
`transverse magnetic field' in units of the flux quantum~\cite{NELSON}.

\begin{figure}
\centerline{\epsfysize=3cm \epsfbox{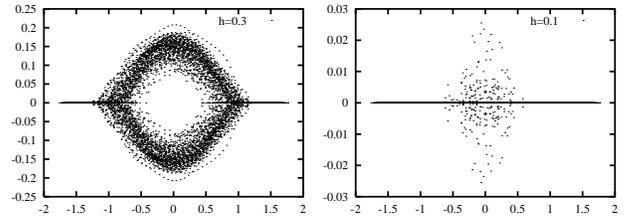}}
\caption{Re $\epsilon$ -- Im $\epsilon$ for $t=\Delta=1$.}
\label{fig-num}
\end{figure}
The eigenvalues of $H$ are complex valued for $h\neq 0$. 
Since $H(-h) = H^T (h)$, the left-eigenvalues are equal to the
right-eigenvalues. Their distribution in the z-plane is shown in
Fig.~\ref{fig-num}, for 
$h=0.1, 0.3$ and $\Delta=t=1$. The results are for an ensemble of 100
matrices  of size $100\times 100$. 
 For $h=0.1$ the eigenvalues are mostly real 
(localized). For $h=0.3$ the eigenvalues around $z=0$ are mostly complex
(delocalized). For larger size matrices, the width of the rim shrinks to zero
in agreement with the results discussed in~\cite{NELSON}.

An important question regarding the character of the spectrum is the
occurrence of a critical $h_c^{(1)}$ for which a gap near $z\sim 0$ sets in 
for the localized states. For increasing $h>h_c^{(1)}$ the eigenvalues migrate
from the real axis to the complex plane as also discussed by Feinberg
and Zee~\cite{FEINBERG}. The migration is total for $h=h_c^{(2)}>h_c^{(1)}$.
To try to quantify this and the bulk 
aspects of the spectrum in Fig.~\ref{fig-num}, we will
analyze~(\ref{master}) using the addition law for free random 
variables~\cite{VOICULESCU,SPEICHER}.

\begin{figure}[h]
\centerline{\epsfysize=17mm \epsfbox{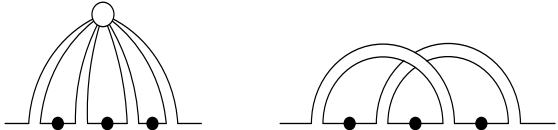}}
\caption{Planar and nonplanar contributions to $\Sigma (z)$. The dotted
lines represent $\GG(z)$.}
\label{fig-diag}
\end{figure}

\vskip .2cm
{\bf 3.\,\,\,}
In this section, we proceed to motivate the use of free random variables
for the Hamiltonian~(\ref{master}). 
For that, we note that
the diagrammatic expansion for~(\ref{master}) follows that of random 
matrix models~\cite{USBIGPAP} to the exception that the $\VV$-propagators 
are changed to
\be
\left< \VV_b^a \VV_d^c \right> =\delta_d^a\delta_b^c \delta_{bd} 
\ee
with a uniform instead of Gaussian measure. Now, consider the resolvent
$G(z)=\tr\GG(z)$ for this model 
\eq
\label{e.green}
\GG(z)=\cor{\f{1}{z-\DD-\VV}}=\f{1}{z-\DD-\Sg(z)}
\eqx
where $\Sg(z)$ is the 1-particle irreducible (1PI) self-energy.
Typical contributions to $\Sigma (z)$ are shown in Fig.~\ref{fig-diag}.
They may be decomposed into planar connected ($pl,\,c$) and nonplanar 
1PI ones ($np$). Generically,
\be
\Sg(z)= \sum_n && (\cor{\tr\VV\GG(z)\VV\ldots
\GG(z)\VV}_{pl,\, c.}
\nonumber \\&&+
\cor{\tr\VV\GG(z)\VV\ldots
\GG(z)\VV}_{np})
\ee
The planar piece is equal to
\eq
\sum_{i=1}^N \cor{V_i^n}_{pl,\, c}\cdot (\GG(z)_{ii})^{n-1}=
\cor{\tr\VV^n}_{pl,\, c} G(z)^{n-1}
\eqx
In the last equality we made use of the invariance of $\DD$
under lattice translations. The nonplanar piece is involved. Here, we
will {\it approximate} ${\cal G}(z)$ it by its {\it diagonal} part
$G(z)\cdot \One$. 
The latter may be moved out of the trace, resulting into
\eq
\Sg(z)=\sum_n \left( \cor{\tr \VV^n}_{pl,\, c}
+\cor{\tr \VV^n}_{np} \right) \cdot G(z)^{n-1}
\label{main}
\eqx
This is our main approximation. 

Having said this, we note that the
calculation of the resolvent would be simplified considerably if we
could show that there {\it exists} a random matrix model $\MM $ with
a measure $P (\MM )$ such that
\eq
\label{e.equ}
\cor{\tr\MM^n}_{pl,\, c}= \left( \cor{\tr\VV^n}_{pl,\, c}
+\cor{\tr\VV^n}_{np} \right)
\eqx
Indeed, then $\Sigma (z)$ and hence $G(z)$ for our problem would simply
follow from the {\it analogue} random matrix model $H_0+\MM$. The main
advantage is that now only {\em planar} graphs in the random matrix model 
contribute, for which the addition law  of free random variables~\cite{ZEE} 
apply. 

To show this, we note that all the nonplanar parts
$\cor{\tr\VV^i}_{np}$ can be absorbed into  effective
connected $i$-th moments. Hence, we may use the combinatorial relations
discussed in~\cite{ITZYKSON}, between connected and ordinary moments for 
{\em planar} diagrams to reduce the equality~(\ref{e.equ}) just to the
equality of moments
\eq
\cor{\tr\MM^n}_{pl}=\cor{\tr\VV^n}
\eqx
with the pertinent measures for each averaging.
This is equivalent to finding a probability distribution
$P(\MM)=\exp(-N\tr {\bf V}(\MM))$ such
that the resolvents of $\MM$ and $\VV$ coincide. To this
end we may solve the loop equation (technically with a single cut) 
of~\cite{AJM} to obtain
\eq
\label{e.measure}
\f{d{\bf V}(z)}{dz}=G_{\VV}(z+i\epsilon)+G_{\VV}(z-i\epsilon)
\eqx
A matrix model with such a potential will satisfy all our
requirements. However to apply the methods of free random variables we
need only to know that such a model exists, the explicit form of ${\bf V}$
through~(\ref{e.measure}) is unnecessary. This concludes our proof of 
existence. 

To summarize, our approximation scheme for~(\ref{master}) in the form
of~(\ref{main}) includes all the planar graphs and also resums an
infinite  
series of (diagonal) nonplanar contributions for the resolvent. It is easily
implemented in the matrix analogue using the addition law for $H_0+{\cal M}$.
Although this procedure is not exact in large $N$, the results
are close to the exact results for large $z$ and large $N$, as
shown in the Table for the first few moments in the Laurent
expansion of the resolvent, that is
\be
G(z) =\sum_{n=0}^{\infty} \,{\bf C}_{2n} \, \frac 1{z^{2n+1}}
\label{laurent}
\ee
for $h=0$, $t=2$ and $\Delta=1$.
The deviations from freeness start at the eight moment and are very small. 
\begin{center}
\begin{tabular}{|c|@{\ }r|@{\ }r|@{\hspace*{10mm}}|c|@{\ }r|@{\ }r|} \hline
$2n$ & ${\bf C}_{2n}^{exact}$ & ${\bf C}_{2n}$ & $2n$ &
  ${\bf C}_{2n}^{exact}$ & ${\bf C}_{2n}$ \\ \hline
2 & 3 & 3 & 	 8 & 736 & 732 \\
4 &16 & 16 & 	10 & 5644 & 5534 \\
6 & 103 & 103 & 12 & 45634 & 43654 \\
\hline 
\end{tabular}
\end{center}

\vskip 0.2cm
{\bf 4. \,\,} 
In this section we derive the resolvent $G(z)$ along the real axis
using the addition law for `hermitean' matrices. $\pi\nu (\lambda) =
{\rm Im} G(\lambda +i\epsilon )$ is the distribution of localized
eigenvalues. With this in mind, the resolvent for the random part is 
simply given by 
\be
G_{\VV} (z) = 
\frac 1N \frac 1{2\Delta} \int_{-\Delta}^{\Delta} \frac{N dV}{z-V}
	= \frac 1{2\Delta} \mbox{ln}\ \frac{z+\Delta}{z-\Delta}
\ee
with an inverse (Blue's function) given by 
$B_{\VV} = \Delta \coth{\Delta z}$, in agreement with~\cite{BZEARLY}. 
The resolvent for the deterministic part is
\be
G_{H_0}(z) = \frac 1N \sum_{n=1}^N\ 
	\frac 1{z-t \cos{(2\pi n/N+ih )}}
\ee
In the large $N$
limit $G_{H_0} (z) = 1/\sqrt{z^2-t^2}$ for $z$ outside the ellipse defined
by
\be
\left( \frac x{\cosh{h}} \right)^2 + \left( \frac y{\sinh{h}}
	\right)^2 = t^2\, .
\label{ellipse}
\ee
and zero inside. The inverse of the resolvent (Blue's function) is just
$B_{H_0} = \sqrt{1/{z^2} + t^2}\,$.


For large $z$ and in the planar approximation (matrix model), the
resolvent $G(z)$
for~(\ref{master}) along the real axis follows from its functional  
inverse $B[G(z)]=z$ through 
the `hermitean' version of the addition law~\cite{ZEE},
 $B=B_{H_0}+B_{\VV}-1/z$. Specifically
\be
z = \sqrt{\frac 1{G^2}+t^2} + \Delta\coth{\Delta G} - \frac 1G
\label{res}
\ee
and is $h$-independent. Along the real axis, the end-points $z_e$
of the spectrum satisfy $z_e\!\equiv\!B(z_c)$, with 
$dB(z_c)/dz\!=\!0$, that is
\be
-\frac 1{z^3} 
	\frac 1{\sqrt{\frac 1{z^2} + t^2}}
	-\frac{\Delta^2}{\sinh^2{\Delta z}} + \frac 1{z^2} =0
\label{dbdz}
\ee
For $\Delta\!\!=t\!\!=\!\!1$ this yields 
$z_c\!\!=\!\!1.5752$ or $z_e\!\!=\!\!1.63915$.

\vskip 0.2cm
{\bf 5. \,\,}
In this section, we derive analytical conditions for the end-points
of Fig.~\ref{fig-num}. For that, we analytically continue $G(z)$ to the 
z-plane minus the ellipse~(\ref{ellipse}), and use it to construct the 
potential $F(z)\!=\!\log Z/N$, where $Z(z)\!=\!\left<{\rm det} (z-H)\right>$. 
In the planar approximation (matrix model) and to leading order in $1/N$, the 
potential follows from $G(z)$ through $F\!=\!\int\!dz\,\, G$,~\cite{USBIGPAP}, 
that is
\be
 F(z)  =&&+
	\Delta G\coth{\Delta G} -1 \nonumber\\ &&+ \log{\frac{1+\sqrt{1+G^2}}{2}}
	-\log{\frac{\sinh{\Delta G}}{\Delta}} \,. 
\label{freout}
\ee
The normalization $Z \sim z^N$ as $z\to\infty$ corresponds to $G(z) \to 1/z$.
For $z\sim 0$, $Z(z)= t/2 e^{Nh} (1 + {\cal O} (N^0))$. This behavior is
related to the other branch ($G (z)=0$) of the resolvent, for which
$F(z)$ with $z\sim 0$ is constant and equal to $h-\log{2/t}$. Hence, a
`cusp' develops in $F(z)$ where the {\it in} and
{\it out} potentials cross, that is
\be
F[G_{out}(z)] =h -\log{2/t} \equiv F[G_{in}(z)] \,.
\label{fene}
\ee
The locus of the cusp~(\ref{fene}) coincides with the position of the
{\em complex} eigenvalue distribution shown in Fig.~\ref{fig-num} as
$N\rightarrow\infty$. The real (localized) eigenvalues are the remnants 
of the hermitean addition law discussed above, and disappear
at some critical value of $h=h_c^{(2)}$.

\begin{figure}
\centerline{\epsfysize=3cm \epsfbox{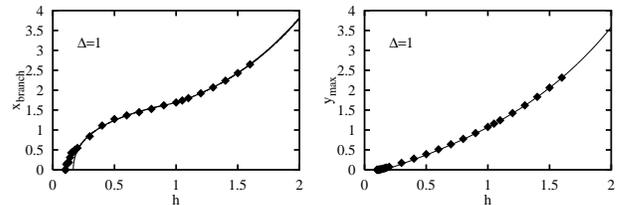}}
\caption{The endpoints of the complex eigenvalue spectra versus $h$ 
at $\Delta\!=\!1$, $t\!=\!1$. See text.}
\label{fig-endp}
\end{figure}

\begin{figure}[tbhp]
\centerline{\epsfysize=3.7cm \epsfbox{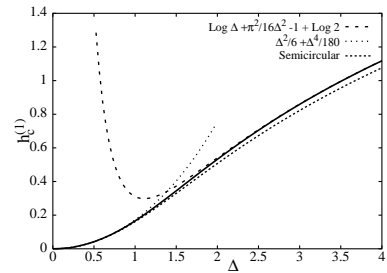}}
\caption{$h_c^{(1)}$ versus $\Delta$ numerically (solid line). For the
analytical curves (dotted and dashed) see text.}
\label{fig-delt}
\end{figure}

Fig.~\ref{fig-endp}a shows the behavior of the critical cusp-line along the
real axis, where the eigenvalue spectra have a branching point
($x_{branch}$), versus  
the strength $h$ for which~(\ref{fene}) holds. At
$x_{branch}(h_c^{(2)})\!=\!z_e$ the localized states disappear from the
spectrum. The dots are the numerically
generated branching points, while the solid line corresponds to using 
(\ref{freout})-(\ref{fene}). Fig.~\ref{fig-endp}b shows the
same along the imaginary axis for the topmost point of the spectrum $y_{max}$. 
For large $h$, the resolvent drops to zero in the outer region like
$1/z$, and using~(\ref{res}) together with~(\ref{freout}) we get for
$y_{max}$ 
\be
y_{max} \approx \sinh{h}-\frac{\Delta^2}{6\sinh{h}}\,.
\ee
The existence of the solution for given $h$ defines the {\it critical} value 
$h_c^{(1)}$ for which the cut starts to develop in the
inner ellipse of Fig.~\ref{fig-num}. The dependence of $h_c^{(1)}$ on $\Delta$
is shown in Fig.~\ref{fig-delt} (solid line). The dotted lines are generated
analytically by working out the leading contributions to $F$ in~(\ref{freout}).
For small values of $\Delta$ (dotted line)
\be
F &=& \frac{\Delta^2}{6}+\frac{\Delta^4}{180}-\log{2}-\frac{i\pi}{2}\,,
\label{fres}
\ee
while for large values of $\Delta$ (long dashed line)
\be
F &=& \log{\Delta}+\frac{\pi^2}{16\Delta^2}-1-\frac{i\pi}{2}\,.
\label{frel}
\ee
\vspace*{-6mm}
These results are also reproduced analytically using a semi-circular 
distribution for the random part, that is
\be
P(V_i) = \frac 1{2\tau\pi}\sqrt{4\tau -V_i^2}
\label{semi}
\ee
when $\tau=\Delta^2/3\rightarrow \infty$, as indicated by the thick
dashed line in Fig.~\ref{fig-delt}.

Finally, the dependence of the eigenvalue distribution shown in 
Fig.~\ref{fig-num} on $1/N$,
maybe qualitatively understood by using the circular
version of~(\ref{master}). This amounts 
to trading $t/2 e^{h}\!\to\!r$ and $t/2 e^{-h}\!\to\!0$.
For a uniform distribution of eigenvalues, the secular equation is
\be
  \prod_i (\lambda-V_i) = -(-r)^N \,.
\label{sec-ran}
\ee
For small values of $\Delta$, hence $V_i$,~(\ref{sec-ran}) can
be solved perturbatively,    
\be
  \lambda^N - \left(\sum_i V_i\right) \lambda^{N-1} = -(-r)^N \,,
\ee
with the ansatz $\lambda_j=r e^{i 2\pi j/N}+\epsilon_j$, giving
$\epsilon_j = - \sum_i V_i/N$. Typically,
$  \langle N^2\epsilon_j^2\rangle = 
	N \langle V_i^2\rangle = N {\Delta^2}/3$, 
showing the corrections to be $\epsilon \sim 1/\sqrt{N}$ in large $N$.

\begin{figure}
\centerline{\epsfysize=3cm \epsfbox{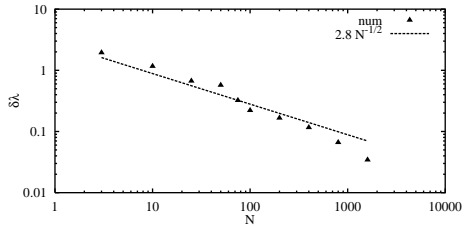}}
\caption{$\delta\lambda$ versus $N$ (matrix size) for $r\!=\!\Delta\!=\!1$.}
\label{fig-wid}
\end{figure}

Numerically the width of the boundary between the outer
($G(z)$ holomorphic) and the inner ($G(z)=0$) maybe analyzed using
\be
  \delta\lambda = \max{(|\lambda_i|-|\lambda_j|)}
\ee
along the real axis. The dependence on the size of the matrix is 
indeed $\sim 1/\sqrt{N}$ as shown in Fig.~\ref{fig-wid}.
\vspace*{-3mm}

\vskip .2cm
{\bf 6.\,\,\,}
The previous analysis borrows on some of 
the methods discussed in~\cite{USBIGPAP} for random matrix models. 
However, the Hatano-Nelson model differs
in an important way from matrix models: it knows about the dimensionality of
space. Recently Efetov~\cite{EFETOV} has used supersymmetric methods to argue
that in 0-dimension the model reduces to a nonhermitean random matrix model 
with weak-nonhermiticity~\cite{FYODOROV}, and suggested that the reduction
may yield to new developments in the context of oriented quantum chaos.

The reduction is actually understandable from the point of view of 
continuum quantum mechanics of constant modes. The continuum 
version of~(\ref{master}) is $H= (p+ih)^2 + {\cal V}$, 
where $p$ is the D-dimensional momentum and ${\cal V}$ the random site potential. 
The reduction to 0-dimension means that ${\cal V}$ is
x-independent. Classically
this would imply that $p=0$. Quantum mechanically, however, $p$ fluctuates.
In a box of size $N$, $p\sim {\cal W}/\sqrt[D]{N}$ by the uncertainty
principle. Hence, the reduction of $H$ to $H_*$ in 0-dimension amounts to 
\be
H_*= {\cal V}  + \frac {i2h}{\sqrt[D]{N}} \, {\cal W}
\label{Q2}
\ee
For sufficiently random hopping, the result is a random matrix model
with small 
nonhermiticity in general. For D=2 this is just the case of {\it weak} 
non-hermiticity discussed by Fyodorov et al.~\cite{FYODOROV}. For ${\cal V,W}$ 
chosen in the GOE ensemble as motivated by ${\cal V}$ real in 
the D-dimensional version of~(\ref{master}),
our arguments suggest localization for D=2 in~(\ref{Q2}) as also 
noted in~\cite{EFETOV}. 

The present arguments may also extend to other models. For example,
the (massless) QCD Dirac operator in a D-dimensional Euclidean box of 
volume $N$ at finite chemical potential $\mu$ is
$H= \gamma^{D+1} (i\gamma^a\nabla^a + i\mu \gamma^D )$, where $\gamma$'s
are Dirac matrices with $a=1,...,D$, and $\nabla$ the covariant derivatives 
with external gauge fields. The squared operator,
\be
H_* = (i\nabla^a)(i\nabla^a) +  \f{i}{2} \sigma^{ab}[i\nabla^a,i\nabla^b]
+2i \mu (i\nabla^D)
\label{Q4}
\ee
with $\sigma^{ab}=i[\gamma^a,\gamma^b]/2$,
is analogous to~(\ref{Q2}). For sufficiently random hopping and
$i\nabla^D\sim 1/\sqrt[D]{N}$, localization may take place for the 
GOE ensemble in D=2 dimensions at finite $\mu$. Since two-dimensional
QCD, in the limit of a large number of colors, 
exhibits quasi long-range order, this issue is worth investigating.

\vskip .2cm
{\bf 7.\,\,\,}
We have shown that the Hatano-Nelson model~(\ref{master}), may be
understood in terms
of the addition law for free random variables. We used the latter to find an 
analytical condition for the end-points of the complex eigenvalue 
distributions. The results compare well with the numerics. The size of the 
spread of the complex eigenvalue distribution is $1/\sqrt{N}$. 
We have presented generic arguments for how the 
constant modes of the model in D-dimensions and for sufficiently random 
hopping, relate to random matrix models with weak nonhermiticity. For D=2
our arguments confirm a recent observation by Efetov~\cite{EFETOV}. In
light of this, it would be interesting to repeat our analysis for $D=2$
for the three Wigner's ensembles.

\vskip .2cm
{\bf Acknowledgments}
\vskip 0.1cm
We would like to thank Prof. D. Nelson for suggesting that we look
at this problem, E. Gudowska-Nowak for discussions. GP thanks
G. Gy\"orgyi, and IZ thanks Y. Fyodorov and A. Zee for discussions. RJ
thanks the Nuclear Theory Group at Stony Brook, where part of this
work was done.
This work was supported in part  by the US DOE grant DE-FG-88ER40388,
by the Polish Government Project (KBN)  grants  2P03B04412  and
2P03B08308 and by the Hungarian Research Foundation (OTKA) grants
T022931 and F019689.  

\setlength{\baselineskip}{15pt}


\vspace*{-5mm}
\begin{thebibliography}{50}
\vspace*{-15mm}
\bibitem{NELSON}
N. Hatano and D.R. Nelson,
	Phys. Rev. Lett. {\bf 77} (1996) 570;
\bibitem{FYODOROV}
Y.V. Fyodorov, B.A. Khoruzhenko, H.-J. Sommers,
	e-print cond-mat/9703152, and references therein.
\bibitem{EFETOV}
K.B. Efetov,
	e-print cond-mat/9702091.
\bibitem{FEINBERG}
J. Feinberg and A. Zee, 
	e-print cond-mat/9703087. 
\bibitem{VOICULESCU}
D.V. Voiculescu,
        Invent. Math. {\bf 104} (1991) 201;
\bibitem{SPEICHER}
  P. Neu and R. Speicher,
	Z. Phys. {\bf B92} (1993) 399;
  P. Neu and R. Speicher, 
	Jour. Stat. Phys. {\bf 80} (1995) 1279.
\bibitem{USBIGPAP}
R. A. Janik, M.A. Nowak, G. Papp and I. Zahed, 
	e-print cond-mat/9612240; and to be published.
\bibitem{ZEE}
A. Zee, 
	Nucl. Phys. {\bf B474} (1996) 726.
\bibitem{ITZYKSON} E. Br\'{e}zin, C. Itzykson, G. Parisi and
J.B. Zuber,
Commun. math. Phys. {\bf 59} (1978) 35.
\bibitem{AJM} J. Ambj{\o}rn, J. Jurkiewicz and Yu. M. Makeenko,
Phys. Lett. {\bf B251} (1990) 517.   
\bibitem{BZEARLY}
E. Br\'{e}zin, S. Hikami and A. Zee, 
	Phys. Rev. {\bf E51} (1995) 5442.
\end{thebibliography}
\end{document}